# Long-term variations in the correlation between NAO and solar activity: the importance of North-South solar activity asymmetry for atmospheric circulation

Katya Georgieva[*]

*Solar-Terrestrial Influences Laboratory at the Bulgarian Academy of Sciences*

Boian Kirov, Peter Tonev, Veneta Guineva,

*Solar-Terrestrial Influences Laboratory at the Bulgarian Academy of Sciences*

Dimitar Atanasov

*Faculty of Mathematics and Informatics, Sofia University*

**Abstract**

General atmospheric circulation is the system of atmospheric motions over the Earth on the scale of the whole globe. Two main types of circulation have been identified: zonal - characterized by low amplitude waves in the troposphere moving quickly from west to east, and meridional with stationary high amplitude waves when the meridional transfer is intensified. The prevailing type of circulation is related to global climate. Based on many years of observations, certain "circulation epochs" have been defined when the same type of circulation prevails for years or decades. Here we study the relation between long-term changes in solar activity and prevailing type of atmospheric circulation, using NAO index reconstructed for the last four centuries as a proxy for large-scale atmospheric circulation. We find that when the southern solar hemisphere is more active, increasing solar activity in the secular solar cycle results in increasing zonality of the circulation, while when the northern solar hemisphere is more active, increasing solar activity increases meridional circulation. In an attempt to explain the observations, we compare the short-term reaction of NAO and NAM indices to different solar drivers: powerful solar flares, high speed solar wind streams, and magnetic clouds.

## 1 Introduction

The problem of solar influences on climate has been discussed for already more than two centuries (Herschel, 1801), and continues attracting increasing attention. Many studies have been published reporting correlations between solar or geomagnetic activity and various climatic parameters. But the results, even when highly statistically significant, are quite contradictory: both positive and negative correlations have been found between solar activity and climatic parameters (see e.g. Herman and Goldberg, 1978 and the references therein). Sazonov and Loginov (1969), based on data since 1884, suggested that the correlation between solar activity and surface air temperature in the 11-year sunspot cycle is negative at low levels of solar activity in the secular solar cycle, and positive in periods of high solar activity. In an earlier study (Georgieva and Kirov, 2000), we compiled all available published results, and found that the correlation between solar activity and surface air temperature in the 11-year sunspot cycle was positive in the whole 18[th] and

[*] Corresponding author. e-mail: kgeorg@bas.bg

20th centuries and negative in the whole 19th century, for both low and high solar activity, and seemed to change systematically in consecutive secular (Gleissberg) solar cycles* (Fig.1).

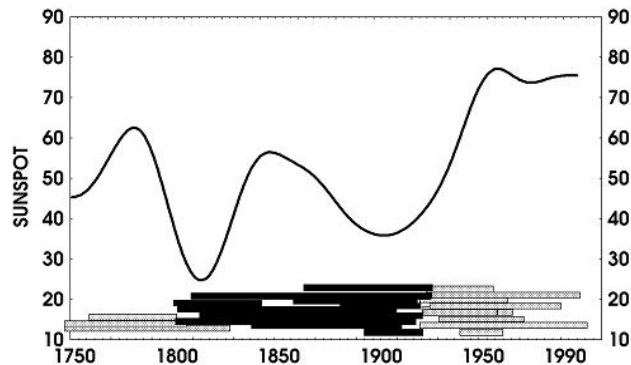

Fig.1. Reported positive (grey bars) and negative (black bars) correlations between sunspot number and surface air temperature in the 11-year sunspot cycle; Solid line denotes sunspot number smoothed by 11-point running average and then 30-point running average (Georgieva and Kirov, 2000).

This result is confirmed by measurements from individual meteorological stations with long data records (Fig.2). The figure was derived in the following way: In the time series of each station, the intervals with length of at least 30 years were identified with the highest statistically significant (positive or negative) correlation between temperature and sunspot number. The whole interval 1700-2000 was divided into 5-year bins, and for each bin the percentage of stations was calculated for which this 5-year period was a part of an interval with statistically significant positive or negative correlation. These numbers were then averaged by a 30-year running mean with a step of 10 years: e.g. 1700-1730, 1710-1740, …1970-2000. Each such 30-year period is represented in Fig.2 by a dark bar whose altitude is proportional to the percentage of stations exhibiting correlations of one sign, and its direction (above or below the x-axis) corresponds to the sign of the correlation – positive or negative, respectively. The sunspot data (solid line in Fig.2) were treated in the same way: the yearly mean values were averaged by a 30-year running mean with a step of 10 years. In meteorology, these averages over 3 full decades are called "normals" and are recommended by the World Meteorological Organization for quantifying climate change (Guttman, 1989). We use this type of averaging in all following figures presenting long-term variations.

Fig.2 demonstrates that all stations which were operating in the 18th century registered positive correlations between surface air temperature and sunspot number in the 11-year solar cycle, in the 19th century in the majority of the stations the correlation was negative, and positive again in the 20th century. Further, we related the change in the correlation between solar activity and temperature to the change in the north-south asymmetry of solar activity, $A=(S_N-S_S)/(S_N+S_S)$ where $S_N$ and $S_S$ stand for the total sunspot area in the northern and southern solar hemispheres, respectively (the white bars in Fig.2): when the southern solar hemisphere is more active, temperature is higher in sunspot minimum and lower in sunspot maximum; and when the northern solar hemisphere is more active, temperature is higher in sunspot maximum and lower in sunspot minimum. If this is true, the secular change in the correlation between solar activity and temperature implies a secular variation in the north-south solar activity asymmetry*.

---

* We should specify that here and in what follows we are studying the correlation of atmospheric parameters with the sunspot number and not with geomagnetic activity. The correlation between sunspot number and aa index is low in both the 11-year solar cycle (Feynman and Gu, 1986) and recently also in the secular solar cycle (Georgieva et al., 2006), however this question is beyond the scope of the present study. A couple of useful references are Lawrence and Ruzmaikin (1998) and Cliver et al. (1998).

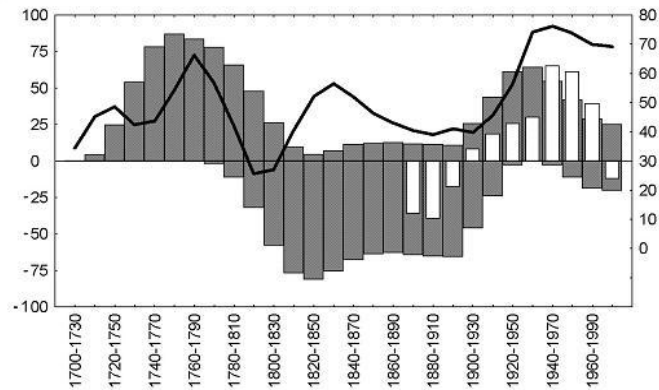

Fig.2. Percentage of meteorological stations (grey bars) measuring positive (along the positive Y-axis) or negative (along the negative Y-axis) correlations in the 11-year sunspot cycle between solar activity and surface air temperature; secular sunspot cycle (solid line); north-south solar activity asymmetry, $A=(S_N-S_S)/(S_N+S_S)$ (white bars) (Georgieva and Kirov, 2000).

Waldmeier (1957) suggested the following relation between the north-south asymmetry of solar activity and the secular solar cycle: solar activity dominates in the northern solar hemisphere during the ascending part of the secular solar cycle, in the southern one during the descending part, and in epochs of secular minima and maxima the asymmetry is small. However, there were at least two episodes when this rule did not hold: the secular maximum of the 20$^{th}$ century around 1950 when the asymmetry had a maximum rather than a minimum (See Fig.2), and the end of the Maunder minimum when the activity was clearly increasing while all sunspots were concentrated in the southern solar hemisphere. From the supposed dependence of the correlation between solar activity and temperature on solar asymmetry (Georgieva, 1998), we can hypothesize that solar asymmetry changes sign in consecutive secular cycles, being positive in "even" cycles (if we denote the 20$^{th}$ century secular cycle as even) and negative in odd ones; it has maximum positive or negative values coinciding with the maximum of solar activity in the odd and even secular cycles, respectively, and changes sign around secular solar minimum. Therefore, we can speak about a "double secular" solar cycle consisting of two secular, or "Gleissberg" (Gleissberg, 1958) cycles – one in which more active is the southern solar hemisphere, and a second one in which more active is the northern hemisphere, much like the 22-year magnetic solar cycle consisting of two 11-year cycles with opposite polarities, which leaves no "anomalies" in the available data (Georgieva and Kirov, 2000). This double secular cycle, inferred from climatic data, was independently determined by Mursula and Zieger (2001) in geomagnetic data: they found that the streamer asymmetry, as determined from seasonal geomagnetic activity, changes its orientation from being shifted towards the southern magnetic heliosphere in the 19$^{th}$ century to being shifted towards the northern magnetic hemisphere in the 20$^{th}$ century.

The reason for the different reaction of surface air temperature to solar activity originating from the two solar hemispheres is not clear, and probably temperature is not the only parameter whose correlation with solar activity depends on solar asymmetry. In the present paper we are dealing with large-scale atmospheric circulation. In Section 2 we study the long-term variations in the correlation between solar activity and atmospheric circulation. In Section 3 we compare the short-term reaction of the circulation to different manifestations of solar activity in the last solar cycle in which a fleet of spacecraft has been measuring solar and interplanetary parameters. We summarize and discuss the results in Section 4.

---

* A useful discussion of the nature of the secular variation and related North-South asymmetry can be found in the book by Zeldovich et al. (1983), where also a possible origin of the asymmetry is indicated (the interplay between the dipole and quadrupole solar magnetic modes) first pointed out by Ivanova and Ruzmaikin (1976).

## 2   Long-term variations in atmospheric circulation

### 2.1  General atmospheric circulation

Atmospheric circulation is the system of atmospheric motions over the Earth on the scale of the whole globe (general atmospheric circulation), or over a certain region with its specific features (local circulation). These large-scale atmospheric motions are caused by the differential heating of the Earth surface. Heated air at the equator rises, creating low pressure there, and proceeds south and north toward the poles. If the Earth didn't rotate, this air would reach the poles, descend and return to the equator, so there would be two circulation cells, one in each hemisphere. However, the Earth does rotate, so when the heated air at the equator rises to a maximum altitude of about 14 kilometers (top of the troposphere) and begins flowing horizontally to the north and south poles, the Coriolis force deflects it, and at latitudes of about 30° the air already flows zonally (parallel to the equator) from west to east. During its motion from the equator, the air cools, and a part of it sinks back to the surface creating the subtropical high pressure zone. Because the Earth's surface is not uniform, the high pressure zone is not a continuous belt but consists of isolated regions of persistent high pressure called "atmospheric centers of action". From this zone, the surface air travels in two directions. One part moves back to the equator completing the circulation system known as the Hadley cell. Another part moves to the poles and is also deflected by the Coriolis force, so at about 60° north and south the surface flow is from west to east. There these subtropical westerlies collide with the cold air coming from the poles, and the collision results in frontal uplift and creation of the "subpolar lows" - the high-latitude centers of action. A small portion of the lifted air flows back to the subtropics after it reaches the top of the troposphere completing the circulation system known as the Ferrel cell. The bigger part is directed to the poles where it creates the polar vortex, moves downward to form the polar high pressure zone, and flows back to midlatitudes in the Polar cell.

The Hadley cell and the Polar cell are "thermally direct" – they exist as a direct consequence of the latitudinal differences in surface temperatures. The Ferrel cell is "thermally indirect" - its existence depends upon the Hadley cell and the Polar cell and comes about as a result of the eddy circulations (the high and low pressure areas) of the midlatitudes. The Hadley and Polar cells are truly closed loops, but the Ferrel cell is not, the prevailing winds in it depend on the passing weather systems. While upper-level winds are essentially westerly (from west to east), surface winds can vary sharply and abruptly in direction. The predominantly westerly winds there can weaken or can become meridional (in the north-south direction), or even easterly for days, changing weather patterns. Therefore, what happens in the Ferrel cells is decisive of the general atmospheric circulation and thus of the terrestrial weather and climate.

### 2.2  Circulation epochs

In 1930's, Vangengeim (1933, 1952) defined three types of atmospheric circulation in the Atlantic-Eurasian region calculated from the daily atmospheric pressure charts over the northern Atlantic-Eurasian region: Westerly (W), Easterly (E), and Meridional (C). Later Girs (1964) applied the same criteria to the circulation in the Pacific-American sector and also found three types of circulation there: one zonal (Z) and two meridional (M1 and M2). W and Z - the zonal types of circulation (westerly at midlatitudes) – are characterized by low amplitude waves in the troposphere moving quickly from west to east. When a meridional circulation type (C or E, M1 or M2) prevails, stationary high amplitude waves are observed and the meridional transfer is intensified. The difference between C and E and between M1 and M2 is in the different location of the waves' crests and troughs. The circulation in the southern hemisphere is less studied, mostly because of the lack of systematic observations over big parts of the oceans, but it was found that many of the features observed in the northern hemisphere are present there too. Again, there are two basic types of circulation – zonal and meridional, the first one divided into two subtypes, and the second one into five (Girs and Kondratovich, 1978).

Girs and Kondratovich (1978) compared the occurrence of the different types of circulation and found that zonal circulation prevailed in the end of the 19$^{th}$ century and in the beginning of the 20$^{th}$ century, then it was replaced by prevailing meridional types of circulation. Their hypothesis was that in periods of long-term increase in sunspot activity the occurrence of meridional forms of circulations is increased, while in periods of decreasing sunspot activity more developed are the zonal forms of circulation. However, the change from prevailing zonal to prevailing meridional circulation coincides with the change in the correlation between sunspot number and surface air temperature in the 11-year sunspot cycle, and with the change in the sign of the north-south solar activity asymmetry, so we can suspect that the decisive factor is not the level or the direction of change of the solar activity, but again the predominantly more active solar hemisphere.

To be able to investigate this possibility, we need longer time series, covering periods of both increasing and decreasing solar activity for more active northern or southern solar hemispheres. Direct data for atmospheric circulation are available only since the end of the 19$^{th}$ century, and for earlier periods proxy data must be used. A large-scale circulation pattern exerting a dominant influence on weather across much of the Northern hemisphere is the North Atlantic Oscillation (Hurrell et al., 2001), and the index of the North Atlantic Oscillation (NAO) representing the large-scale atmospheric circulation in northern midlatitudes, is probably the index with the longest and the most reliable reconstructions (Luterbacher, 1999).

**2.3 Long-term correlation between NAO and solar activity**

The North Atlantic Oscillation is a north-south see-saw oscillation between the atmospheric centers of action in the high-latitude - Icelandic Low, and in the subtropical Atlantic - Azores High, and is defined as the difference between the normalized sea-level pressure in these two centers of action. (The stations used are always Reykjavik in Iceland, and either Gibraltar or Ponta Delgada in the Azores). When the high pressure in the Azores High is even higher than average, usually the low pressure in the Icelandic Low is even lower than average, and NAO is in positive phase characterized by enhanced midlatitude westerly winds across the Atlantic onto Europe. In the reverse case, when the two centers of action are weakened (lower than average high pressure in the Azores High and higher than average low pressure in the Icelandic Low), NAO is negative and zonal circulation is weakened while meridional circulation is enhanced.

The influence of solar activity on NAO in the 11-year sunspot cycle has been studied by a number of authors. Kodera (2003) found that the spatial extend of NAO depends on the level of solar activity: during low solar activity winters, the NAO signal in the sea level pressure is confined in the Atlantic sector, while during high solar activity winters, NAO-related anomalies extend over the northern hemisphere, in particular over the polar region and Eurasian continent. Ruzmaikin and Feynman (2002) showed that the influence of solar activity on the Northern Annual Mode (an analogue to NAO, see below) depends on the phase of the quasi-biennial oscillation (QBO), and that the influence of solar variability on NAM is a part of the possible physical link between solar variability and the low-frequency climate variations (Ruzmaikin et al., 2006).

In a previous study (Kirov and Georgieva, 2002) we used the monthly mean NAO index of Jones et al. (1997) spanning back to 1821. Our conclusion was that the long-term variations of NAO are negatively correlated to sunspot activity: the index has a minimum in the period of the solar activity maximum of the 20$^{th}$ century, and a maximum in the solar activity minimum in the end of the 19$^{th}$ and the beginning of the 20$^{th}$ century with a correlation of –0.73 (Fig.3). The anticorrelation would have been much higher if we had used the data only since the end of the 19$^{th}$ century as earlier the two quantities seem to be actually correlated.

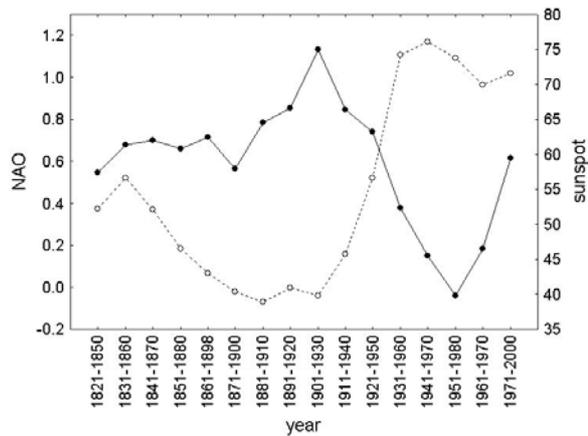

Fig.3. Long-term variations of NAO index (Jones et al., 1997) (solid line), and sunspot number (dotted line): 30-year averages (Kirov and Georgieva, 2002).

A longer NAO reconstruction is now available (Luterbacher et al., 2001) with monthly values since 1659 and seasonal values since 1500. To compare with solar activity, we study the period since 1611 for which the group sunspot number is available as a measure of the yearly level of solar activity (Hoyt and Schatten, 1998). NAO index is calculated as the average of the three winter months, December, January and February. In Fig.4, NAO index is presented (solid line) together with the group sunspot number Rg (dotted line) and the International sunspot number available since 1700 (dashed line). As described in the Introduction, all variables are presented as 30-year averages with a step of 10 years. The correlation between the long-term changes of Jones et al. (1997) reconstruction of NAO presented in Fig.3 and of Luterbacher et al. (2001) reconstruction used in Fig.4 in the period 1821-2000 is 0.87 with $p<0.01$. The vertical lines divide the period into epochs with positive and negative correlations between the long-term variations of NAO and solar activity. (These periods were chosen in such a way as to give the highest positive or negative correlation between NAO and solar activity, so they are somewhat arbitrary). In the $20^{th}$ century the correlation is negative in agreement with our earlier findings, but in the $19^{th}$ century it is positive, negative again in the $18^{th}$ century, and positive again in the $17^{th}$ century. The correlation in the last period is not statistically significant, but it should be noted that this is the period of the Maunder minimum when sunspot activity was atypical. The change in the correlation between the long-term variations of NAO and solar activity coincides with the change in the correlation between surface air temperature and solar activity in the 11-year sunspot cycle, which as noted above, is supposed to coincide with the changes in the long-term solar activity asymmetry. We can therefore speculate that when the southern solar hemisphere is more active, increasing solar activity in the secular solar cycle leads to strengthening of the zonal atmospheric circulation, and when the northern solar hemisphere is more active, increasing solar activity in the secular solar cycle leads to weakening of the zonal circulation.

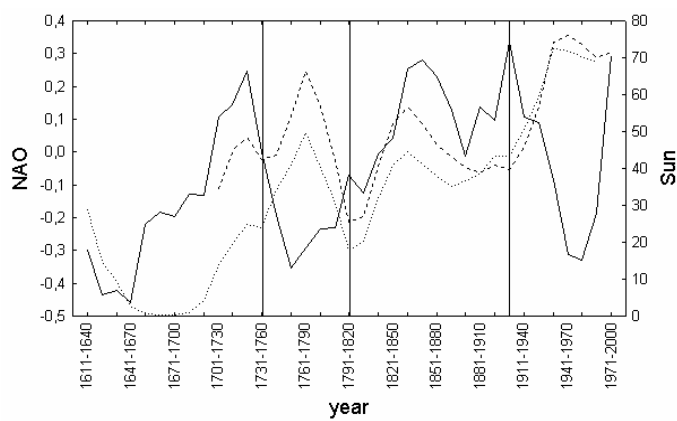

Fig.4. Solid line: NAO index (Luterbacher et al., 2001); dashed line: International sunspot number; dotted line: Group sunspot number (Hoyt and Schatten, 1998). 30-year averages.

## 2.4 Solar activity and the atmospheric centers of action

As noted in Section 2.1, the large-scale atmospheric circulation depends on the position and strength of the atmospheric centers of actions – permanent or semi-permanent areas of persistently high or low pressure. Both observations and models show that these vary in response to solar activity in the 11-year solar cycle. Christoforou and Hameed (1997) studied the solar cycle variations of the semi-permanent Northern Pacific centers of action – the Aleutian Low, active in the cold part of the year (mean latitude 52.5° N and longitude 176° E), and the Hawaiian High, most pronounced in summer (mean latitude 33.5° N and longitude 203°E), and found that during solar maximum conditions Hawaiian High moves northward, and Aleutian Low moves westward and weakens (the area-weighted surface pressure increases).

Haigh et al. (2005) and Haigh and Blackburn (2006) compared observations with a simplified general circulation model results and found a good agreement between the observed atmospheric reaction to the solar cycle and the modeled UV increase: the response to the higher level of solar UV in sunspot maximum consists of weakening and poleward expansion of the Hadley cells and a poleward shift of the Ferrell cells. In other words, energy input into the stratosphere can lead to changes in the tropospheric circulation, even without any forcing below the tropopause. An important result is the dependence of the reaction on the distribution of the stratospheric heating: low-latitude heating forces Hadley cells to move poleward, and high-latitude or latitudinally uniform heating forces them equatorward.

In a previous study (Kirov and Georgieva, 2002) we used the pressure, longitude and latitude monthly averages of the Northern hemisphere centers of action from 1899 to 1995 calculated by Shi (Hameed et al., 1995; Shi, 1999) to study their long-term variations. We found that with increasing solar activity in the secular solar cycle, both subtropical centers of high pressure – Azores High in the Atlantic and Hawaiian High in the Pacific – move poleward and weaken, in agreement with model prediction for weakening and expansion of the Hadley cells in response to increased solar UV (Fig.5).

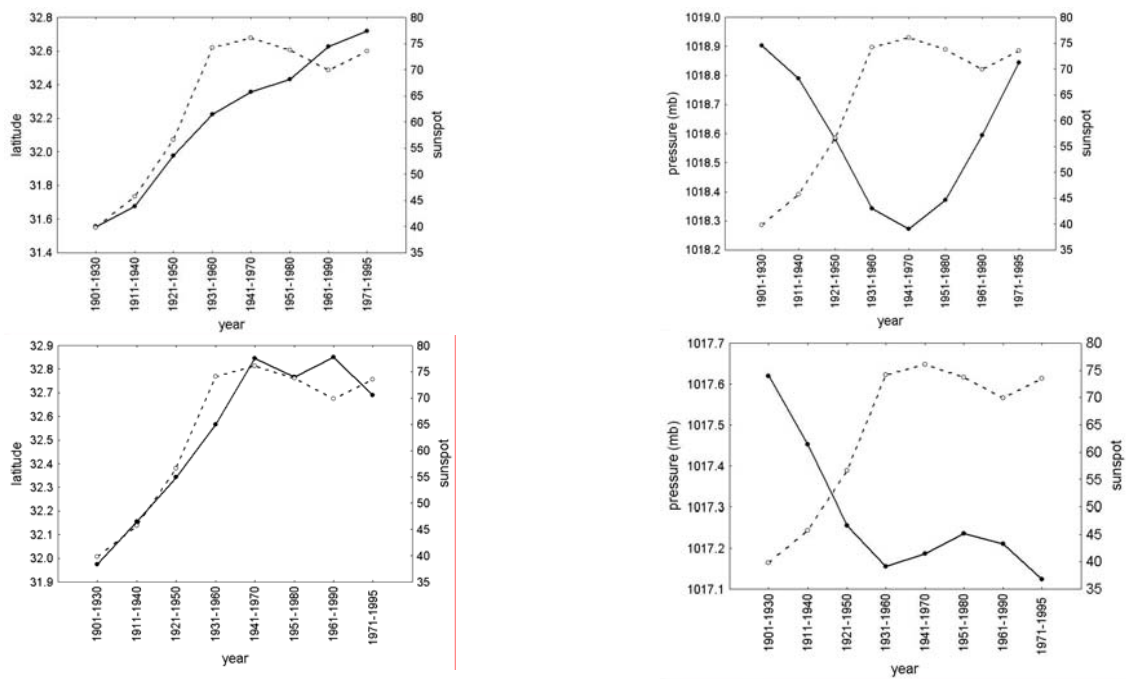

Fig.5. Solid lines: Long-term variations in the latitude (left panels) and pressure (right panels) of the Azores High (upper panels) and Hawaiian High (lower panels); Dashed line: Long-term variations in sunspot number (Kirov and Georgieva, 2005).

As for the mid-latitude centers of low pressure, the Aleutian Low in the Pacific does move poleward in accordance with the expected poleward shift of the Ferrel cells (Fig.6a). However, the latitude of the Icelandic Low in the Atlantic decreases with increasing solar activity in the secular solar cycle (Fig.6b). This means that solar UV is not the main solar activity factor influencing this center of action. According to the model experiments of Haigh et al. (2005), equatorward shifting results from high latitude energy input. The geographic latitude of the Icelandic Low (59.6°N) is not much higher than that of the Aleutian Low (52.5°N), but their geomagnetic latitudes differ much more (Fig.7). Therefore, the Icelandic Low is more subjected to auroral processes related to solar induced geomagnetic disturbances.

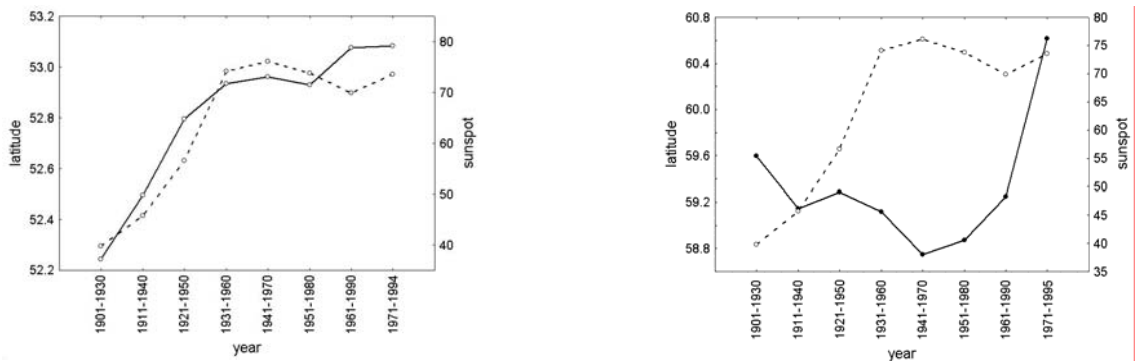

Fig.6. Solid lines: Long-term variations in the latitude of the Aleutian Low (left panel) and Icelandic Low (right panel); Dashed line: Long-term variations in sunspot number (Kirov and Georgieva, 2005).

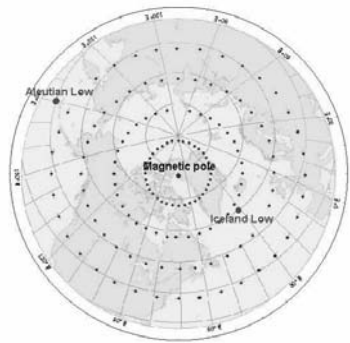

Fig.7. Geographic (solid circles) and geomagnetic (dots) coordinates of Icelandic Low and Aleutian Low

## 3 Short-term reaction of NAO and the atmospheric centers of action to different solar drivers

### 3.1 Events and data

In an attempt to shed some light on the possible reasons for the long-term change in the correlation between atmospheric circulation and solar activity, we study the short-term reaction of the atmosphere to different solar drivers whose number and intensity vary in the course of the 11-year solar cycle, and supposedly also on longer timescales.

The most obvious solar activity agent which can affect the atmosphere and possibly lead to changes in the circulation is solar irradiation, and though total solar irradiation changes by only about 0.1% in the 11-year sunspot cycle (Fröhlich, 2006), the UV radiation which is absorbed in the stratosphere varies by 2-8% (Rottman, 2006). Model results of the effect of increased UV heating in the stratosphere on the tropospheric circulation were briefly described in the previous section, and a mechanism for the downward transmitting of the solar signal is highlighted for example by Kodera (2007). To assess the atmospheric reaction to solar irradiance variations, we study the effects of intense (X-class) solar flares. The list of X-class solar flares used in this study was obtained from the server of the NOAA Space Environment Center http://www.sec.noaa.gov/ftpdir/indices/old_indices/.

The sources of the most intense geomagnetic storms are considered coronal mass ejections (CME's) - eruptions of solar plasma and embedded magnetic field from the corona. (Gonzalez et al., 2002a). In an earlier study we have found that highly geoeffective are not CME's in general but magnetic clouds – a subclass of CME's distinguished by the increased magnetic field intensity, low proton temperature or low plasma beta, and a smooth rotation of the magnetic field over a large angle for a period of the order of a day (Georgieva and Kirov, 2005; Georgieva et al., 2006). This magnetic field rotation, which provides prolonged periods of southward magnetic field and hence geoeffectiveness, is especially interesting for our study, because it is the only characteristic known so far which is persistently different in the two solar hemispheres, irrespective of solar magnetic polarity reversals, and may be the key for understanding the different effects that the two solar hemispheres have on the Earth. The rotating structures on the Sun exhibit clear North-South asymmetry: they are predominantly left-handed in the Northern solar hemisphere and right-handed in the Southern solar hemisphere (Antonucci et al., 1990). On their transit from the Sun to the Earth the magnetic clouds preserve the direction of rotation of the magnetic structures on the Sun from which they originate (Kumar and Rust, 1996), and therefore most of the left-handed magnetic clouds hitting the Earth come from the northern solar hemisphere, while most of the right-handed magnetic clouds originate from the southern solar hemisphere. The list of magnetic clouds which we use was compiled from a number of sources: Fenrich and Luhmann (1998), Leamon et al. (2002), Vilmer et al. (2003), SOHO LASCO CME catalog (http://cdaw.gsfc.nasa.gov/CME_list/), WIND MFI magnetic cloud list (http://sprg.ssl.berkeley.edu/ ~davin/clouds/cloud_list.html), completed by the list of magnetic clouds with identified source regions (Gopalswamy, 2006). The solar wind and IMF parameters for all these events were checked from OMNI database of the National Space Science Data Center http://nssdc.gsfc.nasa.gov/omniweb/, and all events without a clear IMF rotation were excluded.

In a number of papers, C.Jackman and his group (see e.g. Jackman et al., 2006 for a review) presented the effects of solar proton events on the ionization and chemistry in the middle atmosphere, and the subsequent effects on ozone. As the variations in the ozone concentration are related to the thermal regime in the stratosphere and hence to the circulation, we have also studied the reaction of the atmospheric circulation to the proton events. The list of solar proton events was obtained from the Space Environment Center server (http://www.sec.noaa.gov/alerts/SPE.html). It is well known that solar proton events occur during some coronal mass ejections (Kahler et al., 1983), so their effects are expected to be the same as the effects of CME's. Not all magnetic clouds in our list produced solar proton events, and not all solar proton events were associated with magnetic clouds (some were associated with CME's but not magnetic clouds), but even when we removed the overlapping events from both lists, the effects of solar proton events and magnetic clouds were undistinguishable, so we are not showing proton events separately.

The geomagnetic activity during the declining phase of the solar cycle can be even higher that at sunspot maximum. In this period the main drivers of terrestrial disturbances are the high speed solar wind streams originating from solar coronal holes - sources of recurrent geomagnetic storms (Gonzalez et al., 2002b). We have compiled a list of high speed solar wind streams from NASA OMNI data base (http://omniweb.gsfc.nasa.gov/ow.html) defining them as interplanetary structures with high (>500 km/s) speed with a sharp increase in the speed (no less than 100 km/s to no less than 500 km/s in no more that one day) persisting for at least 5 hours, and with high proton temperature and low plasma density. Our definition differs from the one of Lindblad and Lundstedt (1981) by the requirement for high temperature and low density, and from the one used in compiling the ISTP catalog (http://pwg.gsfc.nasa.gov/scripts/sw-cat/Catalog_categories.html) by the requirement for a sharp increase in the flow velocity.

Our catalog of events (magnetic clouds, high speed streams, etc.) is available online at http://stochastics.fmi.uni-sofia.bg/~stil/.

Care was taken not to mix the different types of events. Some magnetic clouds have high speeds, but their interplanetary signatures (low temperature and smooth magnetic field rotation) are very different from the ones of the high speed solar wind streams (high temperature and fluctuating magnetic field components). It is more difficult to distinguish between solar flares and magnetic clouds, because some magnetic clouds are indeed associated with solar flares. Therefore, all cases when a solar flare was followed within 2-5 days by a magnetic cloud, were removed from both lists. Finally, we have 113 cases of high speed solar wind, 103 magnetic clouds, and 66 solar flares in the period 1995-2003. The beginning of the period was determined by the availability of solar and solar wind data (WIND and SOHO satellites), and the end – by the availability of atmospheric data (the project EMULATE).

The project EMULATE (European and North Atlantic daily to MULtidecadal climATE variability) provides the daily gridded sea level pressure data over the extratropical Atlantic and Europe (70°-25° N by 70° W-50° E, 5°x5° boxes, available online at http://www.cru.uea.ac.uk/cru/projects/emulate/). From this data set we use the boxes containing the Azores High (31.6°N, 32.7°W) and the Icelandic Low (59.6° N, 33.1° W), and calculate their difference as a measure of the NAO index. We use this data set rather than the various estimates of NAO in order not only to evaluate the changes in NAO but also to be able to attribute these changes to changes in high or low latitudes.

The Northern Annular Mode is the hemispheric-scale analog of NAO. This is the dominant pattern of non-seasonal sea-level pressure variations north of 20°N, and it is characterized by sea-level pressure anomalies of one sign in the Arctic and anomalies of the opposite sign centered about 37-45°N. In the high phase of the index, the pressure is below normal in the Arctic and the surface westerlies in the north Atlantic are enhanced (Baldwin, 2001). The daily values of NAM at 17 pressure levels from 1000 hPa (surface) to 10 hPa (~32 km) are available online at http://www.nwra.com/resumes/baldwin/nam.php.

### 3.2 Solar flares

Fig.8a presents the changes in the NAO index related to strong solar flares. On the day of the flare the index jumps up and remains at this higher level for 9 days. Fig.8b demonstrates that this increase in the NAO index is due to both reduced pressure in the Icelandic Low, and increased pressure in the Azores High. In this and the following figures, the superposed epoch analysis is employed (Ambroz, 1979): the value on day 0 is the average of all days with events, the values on days -1 and +1 are the averages of all days preceding and following the events, respectively, and so on.

The changes induced by strong solar flares are seen throughout the troposphere. In Fig.9a the NAM-index is plotted at different levels from the surface (1000 hPa) to about 14 km (200 hPa) – the top of the troposphere. The pattern changes above 200 hPa and as we are interested in the troposphere, we will use this boundary as the upper level of our analysis. The two maxima in NAM index (and probably also the prolonged maximum in NAO index) are due to the superposition of two cases mixed in Fig.9a – for positive (Westerly) and negative (Easterly) phases of the Quasibiennial Oscillation (QBO). Fig.9b and 9c demonstrate the importance of the phase of the QBO for the tropospheric reaction to solar induced disturbances: for QBO Westerly (Fig.9b) there is an initial increase in the zonality of the circulation registered at all tropospheric levels on the day of the event, followed by a reduction in zonality, with the net effect being reduced zonality. In QBO Easterly (Fig.9b) the effect is increased zonality seen only at the lowermost levels with a delay of 6 days.

The fact that in QBO Easterly the effects are only seen at lower levels may seem strange as the energy of solar UV is deposited in the stratosphere, but it is in perfect agreement with the results of Haigh and Roscoe (2006) about the combined influence on NAO and NAM of QBO and solar variability as expressed by F10.7 index which is proportional to solar UV.

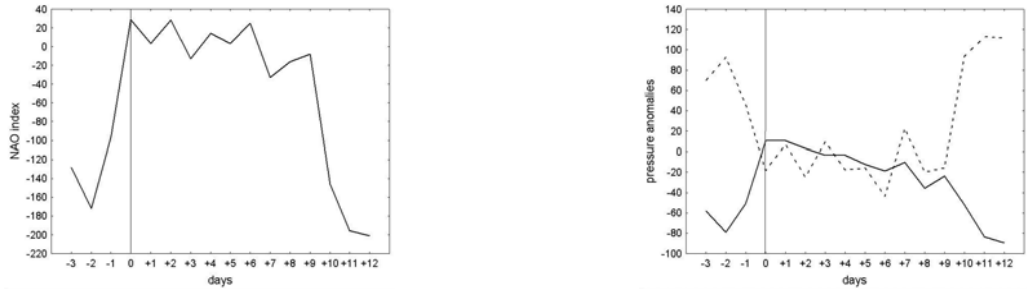

Fig.8. Left: NAO index relative to days with X-class solar flares; Right: surface pressure anomalies in the Azores High (solid line) and Icelandic Low (dashed line) relative to days with X-class solar flares.

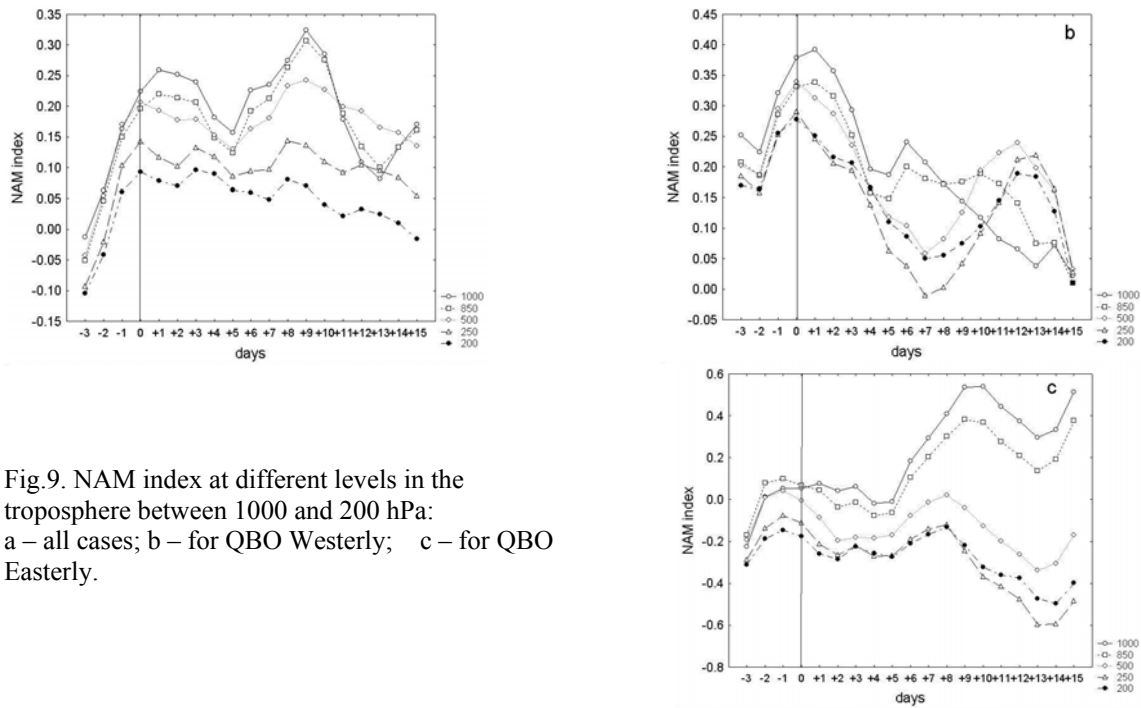

Fig.9. NAM index at different levels in the troposphere between 1000 and 200 hPa:
a – all cases; b – for QBO Westerly; c – for QBO Easterly.

### 3.3 High speed solar wind

The effect of high speed solar wind is an increased pressure impulse with a duration of about 5 days, propagating from high to low latitudes (Fig.10a). This high pressure wave reaches the Azores with a reduced magnitude with a delay of three days, when the pressure in Iceland has already decreased. The net result is an increase in the pressure difference between low and high latitudes and hence in the zonal circulation with a maximum 4 days after the high speed solar wind hits the Earth (Fig.10b).

The influence of the high speed solar wind streams can be seen in the whole troposphere. Fig.11a presents the NAM index at several pressure levels. Everywhere the reaction is increased pressure difference between high and low latitudes, and enhanced zonal circulation. The maximum in NAM index is seen a little later than in NAO index which is natural taking into account the different high and low latitude zones on which the two indices are based.

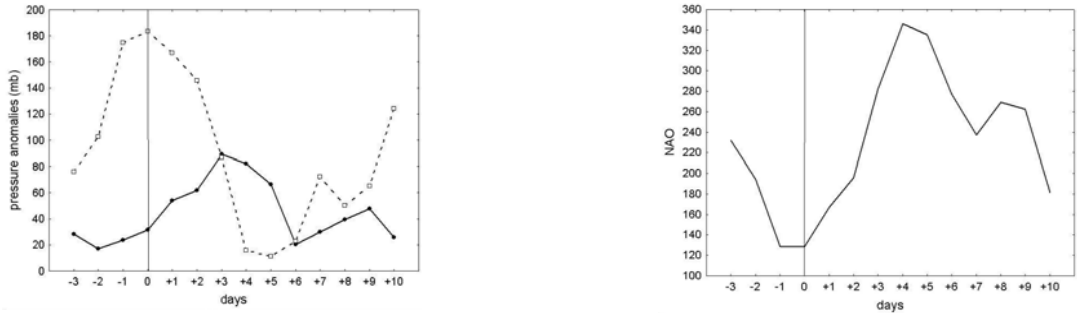

Fig,10. (a) Pressure anomalies in the Azores High (solid line) and Icelandic Low (dashed line) relative to days with high speed solar wind streams; (b) pressure difference between the Azores and Iceland.

The differences between the QBO Westerly and Easterly phases are shown in Fig, 11b and 11c. For QBO Westerly, the maximum in zonality is reached on the seventh day after the high speed solar wind hits the Earth, and NAM index remains elevated for at least one more week. In QBO Easterly, this maximum is on the second day, and by the tenth day NAM index has already returned to its pre-event value.

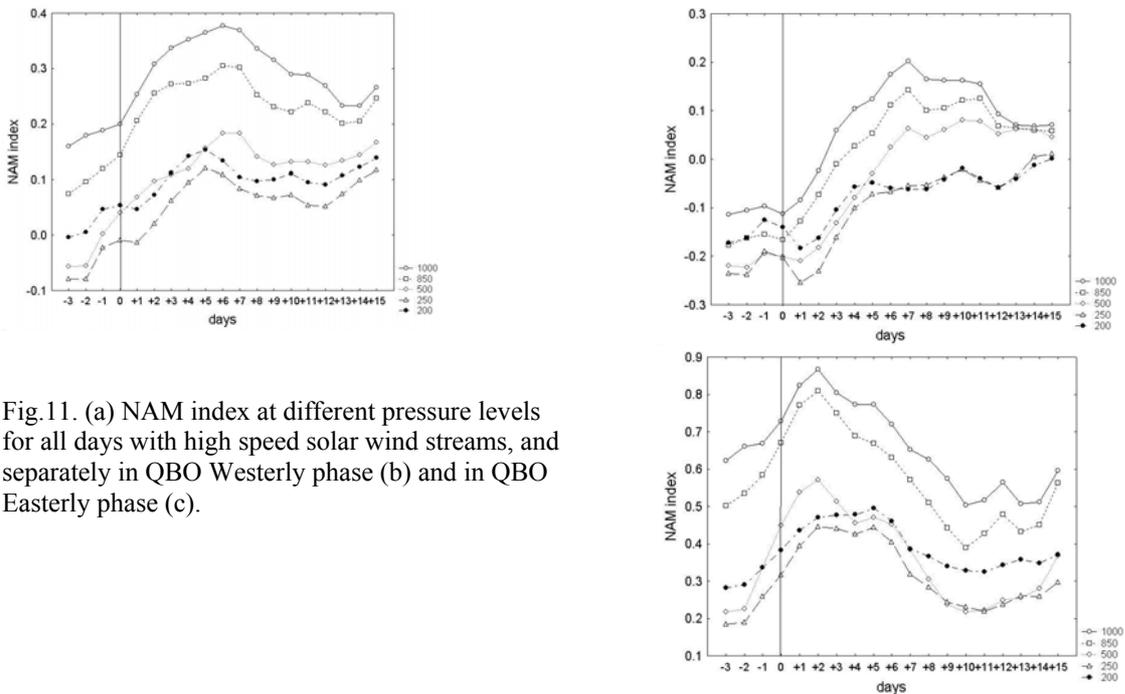

Fig.11. (a) NAM index at different pressure levels for all days with high speed solar wind streams, and separately in QBO Westerly phase (b) and in QBO Easterly phase (c).

### 3.4 Magnetic clouds

Fig.12 presents the reaction of the sea level pressure in the Azores and Iceland, and of their difference as a proxy for zonal atmospheric circulation, for cases of encounter of interplanetary magnetic clouds.

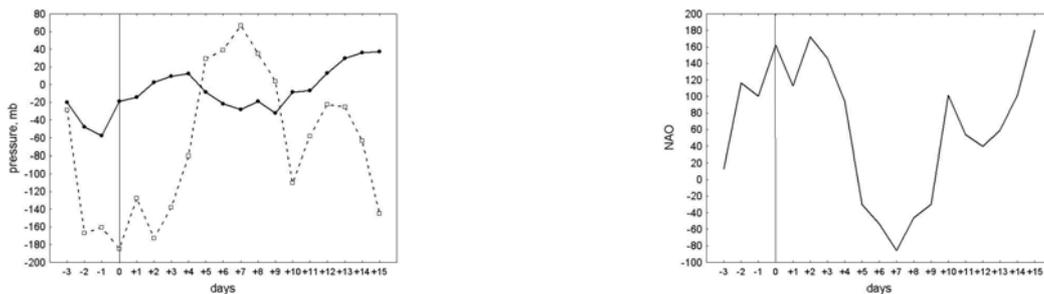

Fig.12. Left: Pressure anomalies in the Azores (solid line) and Iceland (dashed line) relative to days of encounter of magnetic clouds. Right: NAO index as the difference in the pressure between Azores and Iceland relative to the days of encounter of magnetic clouds.

It is obvious that the main reaction is observed at high latitudes, as should be expected taking into account that the effect of magnetic clouds is strongest at high latitudes. The pressure at low latitudes practically doesn't react to magnetic clouds. Consequently, the changes in NAO index are determined by the high latitude processes. NAO has a prolonged deep minimum starting two days after the magnetic cloud encounter, and reaching its lowest values five days later. A more careful inspection reveals that on the day of the encounter of a magnetic cloud and shortly prior to it NAO index has a maximum, and the pressure in Iceland has a minimum. This maximum in NAO and the minimum in Iceland pressure vanish if we exclude the fast magnetic clouds with preceding shocks. Obviously, the shock itself, irrespective of its causes, always leads to a decrease in the pressure at high latitudes. It should be noted here that according to the criteria which we have adopted, the beginning of the magnetic cloud is considered the beginning of the magnetic field rotation, while normally the shock precedes it by a few hours to a day or two.

These changes in atmospheric circulation are best expressed at the surface and close to it (Fig.13a). At higher levels (500 hPa and above) no decrease is seen in NAM index after the encounter of magnetic clouds, and the recovery to normal values observed at lower levels is here manifest as an increase in the index. The main contribution to this picture is of the magnetic clouds encountered in QBO Easterly phase (Fig.13c) whose effect is much bigger (compare the scales of Fig.13b and ,13c) and, as in the case of solar flares (Fig.9c), is confined to low altitudes.

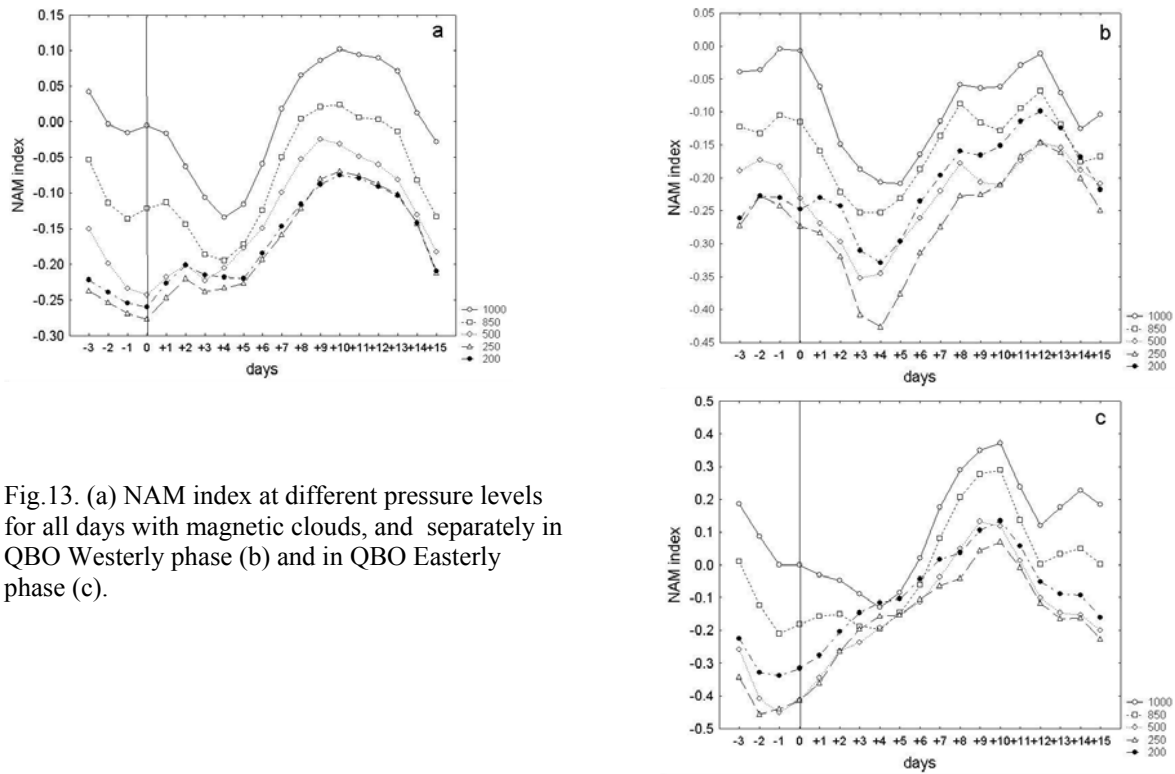

Fig.13. (a) NAM index at different pressure levels for all days with magnetic clouds, and separately in QBO Westerly phase (b) and in QBO Easterly phase (c).

In the case of magnetic clouds, there is one more important parameter which may determine the way in which the atmosphere responds to the influence. As mentioned in Section 3.1, the magnetic clouds differ by their handedness – that is, the direction of rotation of the magnetic field inside the structure which in the majority of the cases corresponds to the direction of rotation of the magnetic structures in the cloud's source region on the Sun. A magnetic cloud is "left-handed" if, viewed from the Earth, the magnetic field in its source region is twisted counterclockwise, and the vector of the magnetic field inside the cloud rotates counterclockwise as measured in either the GSE or the GSM coordinate system. The majority of the left-handed magnetic clouds originate from the solar northern hemisphere. Magnetic clouds with clockwise rotation are "right-handed" clouds, originating predominantly from the solar southern hemisphere.

In Fig.14 the changes in NAM index are compared for cases of right-handed and left-handed magnetic clouds. It is clearly seen that right-handed magnetic clouds have practically no effect on atmospheric circulation at any level, and most of the effects seen in Fig.13a are due to the impact of the left-handed clouds. The influence is more pronounced at lower levels, and consists of a prolonged period of reduced zonal circulation starting a day after the magnetic cloud encounter and reaching its maximum phase some 3-4 days later.

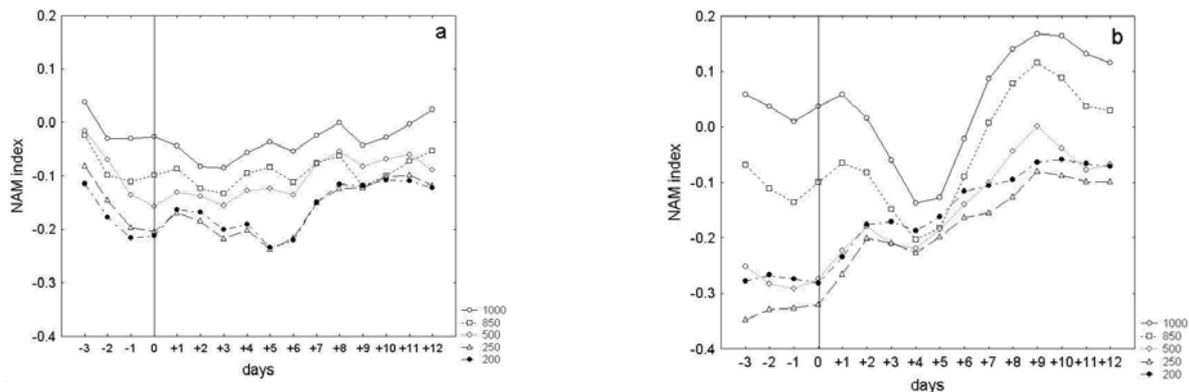

Fig.14. NAM index at different pressure levels for all days with (a) right-handed magnetic clouds, and (b) left handed magnetic clouds.

## 4. Summary and discussion

Using an extended time-series of the NAO index as a proxy for the zonality of atmospheric circulation in the northern midlatitudes, we find that the long-term correlation between solar activity and atmospheric circulation changes in consecutive secular solar cycles and depends on the north-south asymmetry of solar activity: when the northern solar hemisphere is more active, increasing solar activity in the secular (Gleissberg) cycle leads to decreasing prevalence of zonal forms of circulation, while increasing solar activity in secular solar cycles when more active is the southern solar hemisphere leads to increasing zonality of atmospheric circulation.

There are basically two types of solar agents that could affect the Earth and possibly its atmosphere: solar electromagnetic radiation and solar corpuscular radiation. Solar flares are the most powerful manifestations of solar irradiance variations. It is supposed that the strength and/or the number of active regions are proportional to the sunspot cycle amplitude (Wang et al., 2005). Solar flares originate from solar active regions, so the long-term variations in the sunspot cycle amplitude (that is, the secular solar cycle) is proportional to the number and power of solar flares (Mouradian, and Soru-Escaut, 1995). The same is true about solar coronal mass ejections and magnetic clouds as a subclass of coronal mass ejections: both their number and intensity are supposed to be also positively correlated to the secular sunspot cycle (McCracken et al., 2001). As for the high speed solar wind streams, they originate from open magnetic field configurations (solar coronal holes) whose long-term evolution is still a matter of controversy. The long-term decrease in the correlation between sunspot and geomagnetic activity implies different long-term evolution of solar open and closed magnetic field (Georgieva and Kirov, 2006), therefore in long-term studies coronal mass ejections and high speed solar wind should be regarded as different solar drivers rather than generally as solar corpuscular radiation.

The short-term reaction of the atmospheric dynamics to different solar drivers can give some hint about the long-term atmospheric variations in response to solar activity. Solar flares lead to decreased sea level pressure at high latitudes and increased pressure at low latitudes, therefore to increased pressure difference between low and high latitudes and increased zonality of the circulation. The effects at higher altitudes depend on the phase of the Quasibiennial Oscillation of stratospheric winds: in QBO Westerly phase the reaction is immediate and is seen at all tropospheric levels. In QBO Easterly the effect is delayed and is confined to lower levels.

High speed solar wind raises an enhanced pressure wave at the surface propagating from high to low latitudes, with a net result of the superposition of the waves at high and low latitude being an increased pressure gradient and enhanced zonal circulation a few days after the encounter of the high speed stream. The zonality of the circulation is increased in the whole troposphere, and depends on the QBO phase.

The effect of magnetic clouds is increased pressure at high latitude and no effect at low latitude, as a result reduced pressure difference between high and low latitudes and reduced zonal circulation. The effect is the same in the whole tropospheric depth for QBO Westerly, while for QBO Easterly at higher levels an increase rather than a decrease in zonality is observed. The influence of the magnetic clouds also depend on their handedness: it is seen only for left-handed clouds, while right-handed clouds have no effect on tropospheric circulation.

Based on these findings, the following speculations can be made about the long-term dependence of atmospheric circulation on solar activity: At higher solar activity the number and intensity of solar flares is higher, so increasing solar activity leads to increased zonality of atmospheric circulation. This effect does not depend on solar activity asymmetry and should be the same no matter which solar hemisphere is more active.

With increasing solar activity, the number and intensity of magnetic clouds increase, and their effect competes with the effect of solar flares. When the southern solar hemisphere is more active, the magnetic clouds are predominantly right-handed and as right-handed clouds have no influence on tropospheric circulation, therefore the net result is predominantly due to the influence of solar flares and is increasing zonality for increasing solar activity. When the northern solar hemisphere is more active, the magnetic clouds are predominantly left-handed and their effect is a reduction in the zonality of tropospheric circulation. The net result is a superposition of the strengthening of the zonal circulation by solar flares versus its weakening by the magnetic clouds.

It should be reminded here that this study on the effects of magnetic clouds on surface air pressure and atmospheric circulation is confined to the northern hemisphere. It has been found that the long-term variation in the atmospheric circulation in the southern hemisphere is opposite to the one in the northern hemisphere (Georgieva, 2002), and we have some initial results pointing that while the northern hemisphere is only affected by left handed clouds and not by right handed clouds, the reverse seems to be true for the southern hemisphere. However, this requires a more detailed study.

Little can be said about the century-scale variations in the solar open magnetic field and hence of the number and intensity of high speed solar wind streams, but we can suppose that increasing solar open flux would lead to enhanced zonality of the circulation. The decreasing correlation between the long-term variations in sunspot and geomagnetic activity and the recent increase in geomagnetic activity point at the increasing importance of solar open flux. This increase in geomagnetic activity not related to increasing sunspot activity coincides with the gradual change over the past 4 decades of the NAO pattern from the most extreme and persistent negative phase in the 1960's to the most extreme positive phase during the late 1980's and early 1990's – a fact which supports the role of the solar open flux in enhancing the zonality of the circulation.

**Acknowledgements**

This study was partly supported by EOARD grant 063048.